\documentclass[letterpaper,twocolumn,10pt]{article}
\usepackage{usenix}
\usepackage{nopageno}
 
 \usepackage{algorithmic}
 \usepackage{graphicx}
 \usepackage{textcomp}
 \usepackage{comment}
 \usepackage{multirow}
 \usepackage{multicol}
 \usepackage{hhline}
 \usepackage{stfloats}
 \usepackage{color,soul}
 \usepackage{url}
 \usepackage{threeparttable}
 \usepackage{fmtcount}
 \usepackage{xurl}
 \usepackage{hyperref}
 \usepackage{arydshln}
 \usepackage{xspace}
 \usepackage{circledsteps}
 \usepackage{tikz}
 \usetikzlibrary{tikzmark,calc,arrows.meta}
 \usepackage{cprotect}
 \usepackage{enumitem}
 \usepackage[outline]{contour}
 \usepackage{listings}
 \definecolor{codegreen}{rgb}{0,0.6,0}
 \definecolor{codegray}{rgb}{0.5,0.5,0.5}
 \definecolor{codepurple}{rgb}{0.58,0,0.82}
 \definecolor{backcolour}{rgb}{0.95,0.95,0.92}
 \definecolor{darkblue}{rgb}{0.0, 0.0, 0.55}
 \definecolor{darkgreen}{rgb}{0.0, 0.2, 0.13}
 \definecolor{darkred}{rgb}{0.6, 0.0, 0.0}
 \definecolor{darkorange}{rgb}{1.0, 0.27, 0.0}
 \definecolor{lightgreen}{rgb}{0.84, 0.99, 0.82}
 \definecolor{green}{rgb}{0.57, 0.98, 0.48}
 \definecolor{lightred}{rgb}{0.98, 0.90, 0.90}
 \definecolor{red}{rgb}{1, 0.63, 0.63}
 \definecolor{deepred}{rgb}{1, 0.34, 0.34}
 
 \lstdefinestyle{mystyle}{
     backgroundcolor=\color{backcolour},   
     commentstyle=\color{codegreen},
     keywordstyle=\color{magenta},
     numberstyle=\tiny\color{codegray},
     stringstyle=\color{codepurple},
     breakatwhitespace=false,         
     breaklines=true,                 
     captionpos=b,                    
     numbers=left,                    
     showspaces=false,                
     showstringspaces=false,
     showtabs=true,                  
 }
 
 \makeatletter
 \def\namedlabel#1#2{\begingroup
     #2%
     \def\@currentlabel{#2}%
     \phantomsection\label{#1}\endgroup
 }
 \makeatother
 
 \def\checkmark{\tikz\fill[scale=0.2](0,.35) -- (.25,0) -- (1,.7) -- (.25,.15) -- cycle;} 
 
 \def\BibTeX{{\rm B\kern-.05em{\sc i\kern-.025em b}\kern-.08em
     T\kern-.1667em\lower.7ex\hbox{E}\kern-.125emX}}
     
 \usepackage{tcolorbox}
 \newtcolorbox{mtbox}[1]{left=0.25mm, right=0.25mm, top=0.25mm, bottom=0.25mm, colframe=red!50!black, boxrule=0.5pt, title={#1}, fonttitle=\bfseries, coltitle=red!50!black, attach title to upper={\ --\ }}

 \usepackage[framemethod=tikz]{mdframed}
 \mdfdefinestyle{remarkstyle}{
 backgroundcolor=lightgray,
 innerleftmargin=2pt,
 innerrightmargin=2pt,
 innertopmargin=2pt,
 innerbottommargin=2pt,
 linewidth=0.5pt,
 }
 
 \newcounter{summ}[section]

 \newcommand{\sysnamecfi}{{\scshape Sherloc}\xspace}

\begin{document}
       
 \newcommand{\ziming}[1]{%
   \begingroup
   \definecolor{hlcolor}{RGB}{20, 255, 20}\sethlcolor{hlcolor}%
   \textcolor{black}{\hl{\textit{\textbf{Ziming:} #1}}}%
   \endgroup
 }
 
 \newcommand{\jun}[1]{%
   \begingroup
   \definecolor{hlcolor}{RGB}{20, 255, 20}\sethlcolor{hlcolor}%
   \textcolor{black}{\hl{\textit{\textbf{Jun:} #1}}}%
   \endgroup
 }
 
 \newcommand{\gl}[1]{%
   \begingroup
   \definecolor{hlcolor}{RGB}{20, 255, 20}\sethlcolor{hlcolor}%
   \textcolor{black}{\hl{\textit{\textbf{Le:} #1}}}%
   \endgroup
 }
 
 \newcommand{\zheyuan}[1]{%
   \begingroup
   \definecolor{hlcolor}{RGB}{255, 241, 158}\sethlcolor{hlcolor}%
   \textcolor{black}{\hl{\textit{\textbf{Zheyuan:} #1}}}%
   \endgroup
 }
 
 \newcommand{\ZQ}[1]{%
   \begingroup
   \definecolor{hlcolor}{RGB}{255, 20, 20}\sethlcolor{hlcolor}%
   \textcolor{black}{\hl{\textit{\textbf{ZQ:} #1}}}%
   \endgroup
 }
 
 \newcommand{\xt}[1]{%
   \begingroup
   \definecolor{hlcolor}{RGB}{0,32,96}\sethlcolor{hlcolor}%
   \textcolor{white}{\hl{\textit{\textbf{Xi:} #1}}}%
   \endgroup
 }
 
 \newcounter{arch}
 \newcommand{\arch}[1]{
   \stepcounter{arch}
   \namedlabel{#1}{\textbf{A\padzeroes[2]\decimal{arch}}}\textbf{. #1}    
 }
 
 \newcounter{bug}
 \newcommand{\bug}[1]{
   \stepcounter{bug}
   \namedlabel{#1}{\textbf{B\padzeroes[2]\decimal{bug}}}\textbf{. #1}    
 }
 
 \newcounter{limitation}
 \newcommand{\limitation}[1]{
   \stepcounter{limitation}
   \vspace{4pt}\noindent\namedlabel{#1}{\textbf{L\padzeroes[2]\decimal{limitation}}}\textbf{. #1}
 }
 
 \newcounter{issue}
 \newcommand{\issue}[1]{
   \stepcounter{issue}
   \vspace{4pt}\noindent\namedlabel{#1}{\textbf{I\padzeroes[2]\decimal{issue}}}\textbf{. #1}
 }
 
 \newcounter{defense}
 \newcommand{\defense}[1]{
   \stepcounter{defense}
   \vspace{4pt}\noindent\namedlabel{#1}{\textbf{D\padzeroes[2]\decimal{defense}}}\textbf{. #1}    
 }
 
 \newcounter{test}
 \newcommand{\test}[1]{
   \stepcounter{test}
   \textbf{T\padzeroes[2]-\decimal{test}. #1}
 }
 
 \newcounter{recommendation}
 \newcommand{\rec}[1]{
   \stepcounter{recommendation}
   \vspace{4pt}\noindent\namedlabel{#1}{\textbf{R\padzeroes[2]\decimal{recommendation}}}\textbf{. #1} 
 }
 
 \pagestyle{plain}
 \newcommand{\tabincell}[2]{\begin{tabular}{@{}#1@{}}#2\end{tabular}}
 
 \newcommand*{\priority}[1]{\begin{tikzpicture}[scale=0.08]%
 		\draw (0,0) circle (1);
 		\ifthenelse{#1>0}{\fill[fill opacity=1,fill=black] (0,0) -- (90:1) arc (90:90-#1*3.6:1) -- cycle;}{}
 \end{tikzpicture}}
 
 \newcommand*\circled[1]{\tikz[baseline=(char.base)]{
     \node[shape=circle,draw,inner sep=2pt] (char) {#1};}}
 
 \newcommand*\wcircled[1]{\tikz[baseline=(char.base)]{
     \node[shape=circle,draw,inner sep=1pt] (char) {#1};}}
 
 \setlength{\unitlength}{.5em}
 \newcommand\supportlevel[1]{\begin{picture}(1,1)
 \ifnum0=#1\put(.5,.35){\circle{1}}\else
 \ifnum10=#1\put(.5,.35){\circle*{1}}\else
 \put(.5,.35){\circle{1}}\put(.5,.35){\circle*{.#1}}
 \fi\fi\end{picture}}
 
\title{SoK: Where's the ``up''?! A Comprehensive (bottom-up) Study on the Security of Arm Cortex-M Systems}

 \author{
 	{\rm Xi Tan}\\
 	CactiLab, University at Buffalo
 	\and
 	{\rm Zheyuan Ma}\\
 	CactiLab, University at Buffalo
 	\and
 	{\rm Sandro Pinto}\\
 	Universidade do Minho
 	\and
 	{\rm Le Guan}\\
 	University of Georgia
 	\and
 	{\rm Ning Zhang}\\
 	Washington University in St. Louis
 	\and
 	{\rm Jun Xu}\\
 	University of Utah
 	\and
 	{\rm Zhiqiang Lin}\\
 	Ohio State University
 	\and
 	{\rm Hongxin Hu}\\
 	University at Buffalo
 	\and
 	{\rm Ziming Zhao}\\
 	CactiLab, University at Buffalo
 } 
 

\maketitle
 
\begin{abstract}
Arm Cortex-M processors are the most widely used 32-bit microcontrollers among embedded and Internet-of-Things devices. 
Despite the widespread usage, there has been little effort in summarizing their hardware security features, characterizing the limitations and vulnerabilities of their hardware and software stack, and systematizing the research on securing these systems.
 %
The goals and contributions of this paper are multi-fold. 
First, we analyze the hardware security limitations and issues of Cortex-M systems. 
 Second, 
 we conducted a deep study of the software stack designed for Cortex-M and revealed its limitations,
 which is accompanied by an empirical analysis of 1,797 real-world firmware. 
 Third, we categorize the reported bugs in Cortex-M software systems. 
 Finally, we systematize the efforts that aim at securing Cortex-M systems and evaluate them in terms of the protections they offer, run-time performance, required hardware features, etc.
 Based on the insights, we develop a set of recommendations for the research community and MCU software developers.
 
 \end{abstract}
 

 \section{Introduction}


Microcontroller units (MCUs) are small computers designed for embedded and Internet of Things (IoT) applications in contrast to microprocessors used in smartphones, personal computers, and servers. 
They operate at frequencies ranging from several kHz to several hundred MHz.
The sizes of their ROMs and RAMs are small and usually fall into the range of several hundred bytes to several megabytes.
Even though MCUs are general-purpose computers, they are commonly employed for running specialized software and firmware tailored to specific applications.

The Arm Cortex-M family, which has three major architectures and 12 processors as of 2023, is the most popular 32-bit MCU architecture without a memory management unit (MMU) on the market.
More than 80 hardware vendors have licensed Cortex-M cores~\cite{cmip}.
4.4 billion Cortex-M MCUs were shipped in the 4th quarter of 2020 alone~\cite{Armnews2020}, and it is estimated that Cortex-M MCUs account for almost 100 billion deployed embedded and IoT devices in 2021~\cite{ArmBlueprintNews}.

\begin{figure*}[!ht]
	\begin{centering}
		\includegraphics[width=.9\textwidth]{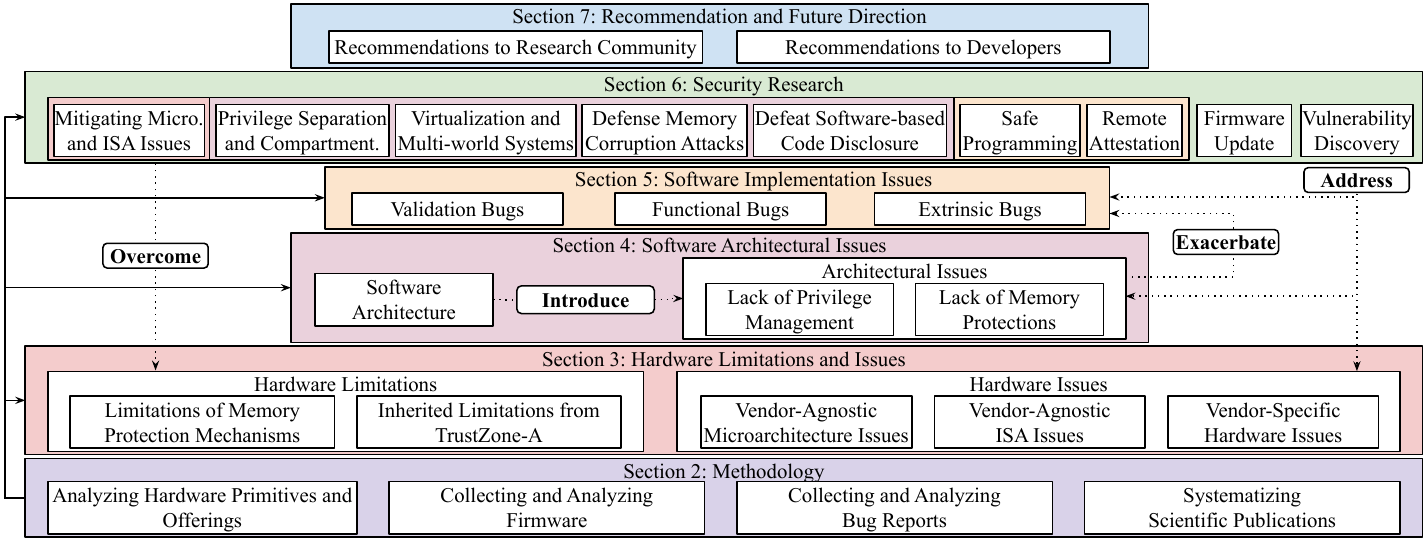}
		\vspace{-0.4cm}
		\caption{Overview of the organization and contributions of this paper}
		\label{fig:sections}
	\end{centering}
	\vspace{-0.6cm}
\end{figure*}

Given the sheer volume of deployed Cortex-M systems, 
one would anticipate that the security of their hardware and software stack has been thoroughly studied and systematized. 
Unfortunately, this is not the case.
To bridge the knowledge gap that hinders the users and researchers,  
we seek to answer the following questions regarding their security states:
\begin{itemize}[noitemsep,topsep=0pt,leftmargin=*]
\setlength\itemsep{0em}
\item \emph{Q1 - What are the security features, limitations, and issues at the Cortex-M microarchitecture, instruction set architecture (ISA), and beyond?} The answer helps understand the constraints in securing software on Cortex-M.

\end{itemize}

To address this question, we analyze the hardware security limitations of Cortex-M by comparing its offerings with microprocessors. 
\emph{Our main observation (\S\ref{sec:HardIssueVul}) is that Cortex-M processors lack support for memory virtualization and provide only basic memory protection mechanisms. Additionally, their other security features, e.g., TrustZone, are streamlined compared to their Cortex-A counterparts and introduce new vulnerabilities.}

\begin{itemize}[noitemsep,topsep=0pt,leftmargin=*]
\setlength\itemsep{0em}
\item \emph{Q2 - What are the security mechanisms and flaws of Cortex-M based software systems?}
The answer helps understand the status of Cortex-M software security in real-world systems. 
\end{itemize}

To answer this question, we compile a dataset of 1,797 real-world Cortex-M firmware samples, including 1,003 newly collected ones, and perform by far the largest empirical analysis on the adoption of security mechanisms on real-world Cortex-M systems.
In particular, we summarize the software architectures found in these samples and other research projects.
We develop binary analysis tools to verify if the collected samples leverage the security mechanisms that have been widely deployed on microprocessor-based systems, e.g., privilege separation and stack canaries.
\emph{We uncovered that (\S\ref{sec:SoftIssueVul}) despite extensive research on more secure architectures for microcontroller-based systems, these advancements are rarely implemented in real-world firmware. Moreover, the hardware security features offered by Cortex-M processors are seldom utilized in the majority of the assessed firmware; hence, where is the ``up''?!. 
Furthermore, existing compiler-based mitigations designed for process-based operating systems (e.g., stack canaries) prove ineffective when operating within a single physical address space}.

\begin{itemize}[noitemsep,topsep=0pt,leftmargin=*]
\setlength\itemsep{0em}
\item \emph{Q3 - What are the nature and severity of the publicly disclosed vulnerabilities in the Cortex-M based software systems?}
The answer helps find out software bugs that are more likely to be exploited in such systems.
\end{itemize}

To tackle this question, we analyze 310 Cortex-M related software bug reports spanning nearly six years, from 2017 until 2023.
Our analysis includes systems developed by nine hardware vendors, e.g., Nordic and NXP, and seven real-time operating systems (RTOS), e.g., FreeRTOS.
We further categorize the software implementation issues into validation, functional, and extrinsic bugs, a taxonomy adopted in a recent work studying the vulnerabilities in Cortex-A systems~\cite{cerdeira2020sok}.
\emph{Our insights (\S\ref{sec:ImpIssueVul}) include that these systems not only exhibit memory corruption vulnerabilities but also display weaknesses in their protocol and cryptographic implementations.} 

\begin{itemize}[noitemsep,topsep=0pt,leftmargin=*]
\setlength\itemsep{0em}
\item \emph{Q4 - What defenses for Cortex-M systems have been explored in the literature, and what are their limitations?} Together with the previous answers, this helps shed light on new research directions to secure Cortex-M systems. 
\end{itemize}

To address this question, we create a taxonomy and comparative evaluation of over 50 papers spanning nearly nine years.
Our evaluation framework considers the defenses each solution offers, the hardware units it relies on, and their run-time overhead in terms of memory size, performance, etc.
\emph{Our major observations (\S\ref{sec:projects}) include the research community not only shifts the exact same defenses from microprocessor-based systems on Cortex-M systems, e.g., enforcing isolation and confinement, stack integrity, and control flow integrity, but also develops solutions  intrinsically linked to the MCU characteristics, e.g., peripheral-oriented fuzzing}.

Based on the insights, we develop a set of recommendations for the research community and MCU software developers (\S\ref{sec:rec}).
Figure~\ref{fig:sections} provides an overview of the organization and contributions of this paper.
We have open-sourced our source code, dataset, and supplementary materials~\footnote{https://github.com/CactiLab/SoK-Cortex-M}.
 
 
 \vspace{-0.2cm}

\section{Methodology}
\label{sec:methodology}

\subsection{Adversarial Model}

In general, we consider the security limitations and issues of the microarchitecture, ISA, and above. 
In particular, we assume an adversary can perform (i) microarchitecture side-channel attacks, e.g., bus interconnect; (ii) glitching, e.g., voltage fault injection; (iii) remote attacks via a network; (iv) nearby wireless attacks via BLE, ZigBee, etc.; (v) local attacks through peripherals and debug ports; and (vi) software side-channel attacks. 
On systems without TrustZone-M, we consider an adversary with one or more of the following objectives: 
(i) to obtain secrets from the flash, e.g., intellectual property (IP) theft and RAM;
(ii) to tamper sensitive data;
(iii) code execution and privilege escalation, e.g., control-flow hijacking.
On systems with TrustZone-M, we assume all components in the non-secure state are untrusted and consider an adversary with all aforementioned goals as well as compromising the secure state.

\subsection{Analyzing Hardware Offerings}

We provide a detailed analysis of the hardware security limitations and issues.
Due to the page limit, a detailed walk-through of the Cortex-M architecture is not included in this paper. 
Interested readers please refer to our supplementary materials, which consolidate information from various official sources~\cite{Armv6MArcRef, Armv7MArcRef, Armv8MArcRef, Armv8MArcTech, Armv8MMPU, ArmCortexM23, ArmCortexM33, ArmCortexM55, Armv8MTZ, ARMpacbti}. 
To aid in research for the community, we have developed an open-source code suite, demonstrating the use of Cortex-M security features. 

\subsection{Collecting and Analyzing Firmware}
\label{sec:analyzing_firmware}

\begin{table}[t]
	\center
	\scriptsize
	\caption{Manufacturer distribution of the compiled real-world firmware dataset. \textit{Italic} represents newly collected sample that were not publicly released before.}
	\label{tab:fws}	
	\renewcommand{\arraystretch}{1}
	\setlength\tabcolsep{0.75ex}
	\rowcolors{2}{gray!15}{white}
	\begin{tabular}{l|c|c|c|c|c|c|c|c|c}
		\hline
		\multicolumn{1}{c|}{HW Vendor} & \begin{tabular}[c]{@{}c@{}}Nordic\\ \cellcolor{white}~\cite{wen2020firmxray}\end{tabular} & \begin{tabular}[c]{@{}c@{}}\textit{Other}\\ \cellcolor{white}\textit{Nordic}\end{tabular} & \begin{tabular}[c]{@{}c@{}}TI\\ \cellcolor{white}~\cite{wen2020firmxray}\end{tabular} & \textit{Telink} & \textit{Dialog} & \textit{NXP} & \textit{Cypress} & \begin{tabular}[c]{@{}c@{}}ST\\ \cellcolor{white}~\cite{gritti2022heapster}\end{tabular} & Total \\ \hline \hline
		\# Firmware                    & 768                                                                                       & \textit{690}                                                                              & 22                                                                                    & \textit{192}    & \textit{53}     & \textit{1}   & \textit{67}      & 4                                                                                        & 1,797 \\ \hline
		\# Devices                     & 513                                                                                       & -                                                                                         & 20                                                                                    & \textit{120}    & \textit{36}     & \textit{1}   & -                & -                                                                                        & 689   \\ \hline
	\end{tabular}
	\vspace{-0.6cm}
\end{table}

\textbf{Collecting Firmware}.
The process of collecting and decoding Cortex-M firmware was far from straightforward and resulted in the accumulation of significant amounts of unusable data.
We used three approaches to collect firmware:
(i) we filtered Cortex-M firmware from publicly available embedded system datasets~\cite{fwheapster, fwgalaxybuds, fwshannon, friebertshauser2021polypyus, fuzzware, wen2020firmxray};
(ii) adopting an analogous methodology as described in~\cite{wen2020firmxray}, we developed scripts to analyze/unpack mobile apps and extract potential Cortex-M firmware. 
Using this approach, we collected 4,693 potential samples from six silicon vendors. 
These samples are in various formats, e.g., S-record for NXP, cyacd format for Cypress, and proprietary format of Qualcomm;
(iii) we crawled websites for 25 silicon and device vendors known for embedded and IoT devices. 
This effort resulted in 1,687 potential samples, but none of them turned out to be Cortex-M firmware. 
This aligns with the findings in FirmXRay~\cite{wen2020firmxray}, which noted that vendors seldom make their firmware available online.

As shown in Table~\ref{tab:fws}, our firmware collection endeavor ended up with 1,797 unique Cortex-M firmware from seven hardware vendors.
Among these, the FirmXRay dataset includes 790 firmware samples, representing 533 distinct devices from two vendors (768 from Nordic~\cite{nordic} and 22 from Texas Instruments~\cite{texasi}). 
Additionally, the HEAPSTER dataset~\cite{gritti2022heapster} encompasses four Cortex-M binaries from STMicroelectronics (ST)~\cite{STM}.
Furthermore, we have gathered 1,003 firmware from other vendors, including Nordic (690), Telink~\cite{telink} (192 firmware for 120 unique devices), Dialog~\cite{dialog} (53 firmware for 36 devices), NXP~\cite{nxp} (1), and Cypress~\cite{cypress} (67). These samples have not been publicly shared before.
The firmware in our collection is in raw binary format, lacking symbolic information.

\textbf{Analyzing Firmware}.
We used FirmXRay~\cite{wen2020firmxray} to recognize the base address of each firmware.
Scripts were then developed to identify the Cortex-M vector table and perform recursive disassembly with Ghidra~\cite{ghidra}.
We also applied scripts to filter a portion of firmware samples that contain device information,  ensuring that they are from distinct devices.
We conducted an analysis of the disassembled samples using the following heuristics:
(i) to identify if firmware uses any RTOS, we performed binary function recognition~\cite{bao2014byteweight} and string searches for ten popular RTOSs;
(ii) for firmware that uses an RTOS, we analyzed if task stack overflow checks are performed.
To this end, 
we checked if the task stack overflow handling functions, e.g., \verb|osRtxKernelErrorNotify()| with the parameter \verb|osRtxErrorStackOverflow| in CMSIS RTOS2~\cite{cmsisrtos2ovf}, are called by other functions in the firmware;
(iii) we analyzed if and how the \verb|CONTROL| register is changed and how the \verb|SVC| instruction is used to determine privilege separation and stack usages;
(iv) to check if there are stack canaries, 
we analyzed function prologues and epilogues for specific instruction patterns derived from canary-protected functions generated by three compilers.
In addition, we searched if the firmware has the hard-coded libc error message ``\verb|*** stack smashing detected ***|'' and whether the function printing out this message is called by other functions, which is a practice used before~\cite{yu2022building}.

\subsection{Collecting and Analyzing Bug Reports}

We retrieved over 500 hardware and software bug reports related to Cortex-M systems from 2017 to 2023~\cite{mitrecve}, which shows a growing trend. 
Besides ``Arm", we included in our list of keywords the names of top hardware vendors~\cite{mcumarkets}, 
popular RTOSs~\cite{rtosmarket}, and embedded SSL libraries, e.g., Mbed TLS~\cite{Mbedtls} and wolfSSL~\cite{wolf}). 
We manually confirmed the bug reports indeed affect Cortex-M systems,
including verifying the affected chips and inspecting the source code. 
Two researchers worked together to categorize each bug into a
relevant subclass, which was verified by a third researcher. 

\subsection{Systematizing Scientific Publications}

We collected over 30 papers on Cortex-M security from top conferences\footnote{https://csrankings.org/}. 
In addition, we supplement our list of surveyed papers with another over 20 articles that are highly relevant to the topic but published in other venues.
Note that our systematization focuses on the works explicitly designed for and implemented on Cortex-M.  
Nevertheless, we discuss related works that were designed for or implemented on other architectures but may be applied to Cortex-M in~\S\ref{sec:discussion}.

\subsection{Threats to Validity}
Our analysis of firmware may be subject to biases and imprecision due to the limited number of firmware. 
There is a risk of over-representing systems from specific vendors.
Most firmware in our dataset (57.3\%) are raw binaries and lack detailed device and architecture information, making it difficult to confirm their intended use cases and resulting in a potential bias in analyzing similar firmware samples.
Additionally, the lack of proof-of-concept exploits and vague CVE descriptions introduces imprecisions in the classification of vulnerabilities. 
Furthermore, our analysis focuses on publicly disclosed vulnerabilities. Undiscovered vulnerabilities could unveil additional fundamental issues in Cortex-M systems.
 
 
 \section{Hardware Limitations and Issues}
\label{sec:HardIssueVul}

\subsection{Hardware Limitations}
\label{subsec:hw_limits}

Hardware limitations are missing or constrained hardware security features, which are typically non-patchable.
Compared with Cortex-A, Cortex-M features distinct design elements, particularly in its memory protection mechanisms and the TrustZone extension (TrustZone-M versus TrustZone-A). 

\vspace{.2cm}\centerline{\ul{Limitations of Memory Protection Mechanisms}}

\limitation{No memory virtualization:} 
No hardware-supported memory virtualization is available on Cortex-M due to the absence of a memory management unit (MMU). 
Instead, software modules share the same physical address space.
Such lack of memory virtualization also implies a small address space (4GB),
which presents challenges to effective address space layout randomization (ASLR) due to low entropy.

\limitation{No input-output memory management unit:}
Besides MMU, input-output memory management unit (IOMMU) or its equivalents, i.e., IOMPU, that provide memory protection from malicious direct memory access (DMA)-capable peripherals are also missing on Cortex-M.
Some hardware vendors implement their own IOMPU, i.e., the resource domain controller on NXP i.MX RT~\cite{nxprt, nxpxrdc},
but they are only found in some of the latest devices.

\limitation{A small number of MPU regions and limited sizes:} 
Cortex-M only supports a small number of memory protection unit (MPU) regions, and the size of regions must be a multiple of 32 bytes.
Compared to the page-based memory access control on microprocessors, the granularity of MPU-based is coarse-grained, and it is insufficient to implement fine-grained isolation that requires a large number of regions. 


\limitation{A small number of secure/non-secure memory regions:}
The number of regions supported by secure attribute unit (SAU) is small, e.g., up to 8 regions on Cortex-M33, 
resulting in limited design choices in splitting the secure and non-secure address space.
To alleviate this issue, silicon vendors use the implementation defined attribution unit (IDAU),which supports up to 256 regions, to create more partitions.
However, if more than 256 partitions are needed or the device has many peripherals, this may not be enough~\cite{sauidaumpcppc}.

\vspace{.2cm}\centerline{\ul{Inherited Limitations from TrustZone-A}}

\limitation{No intrinsic encryption to protect the secure state memory:}
TrustZone-M does not
encrypt the secure state memory.
Consequently, cold boot attacks~\cite{halderman2009lest} can dump the secure state memory.
There could also be information leakage when a memory protection controller (MPC) assigns a memory region from the secure state to the non-secure state at run-time, which we will discuss in \ref{Information leakage to the non-secure state due to state switches:}.

\limitation{Lack of intrinsic support for multiple trusted execution environments:}
TrustZone-M only provides \emph{one} isolated execution environment 
in which the trusted firmware
executes, resulting in a large software trusted computing base (TCB). 
For instance, TF-M~\cite{TFMsrc} has over 117K lines of 
code.

\limitation{Lack of hardware-based remote attestation in TrustZone-M:}
Same as Cortex-A~\cite{cerdeira2020sok}, Cortex-M TrustZone lacks a hardware-based integrity reporting mechanism,
so it cannot provide a hardware-based remote attestation as Intel software guard extensions (SGX) does.
For example, the Arm platform security architecture (PSA) introduces a weakened software-based attestation method~\cite{ArmPSASM, ArmPSAattestation}. 

\begin{mdframed}[style=remarkstyle,backgroundcolor=blue!10]
\centerline{\textbf{Insights}}

\begin{itemize}[noitemsep,topsep=0pt,leftmargin=*]

\item The Cortex-M architecture offers weaker memory management interfaces than popular microprocessors, creating challenges to enforce memory isolation and security.

\item TrustZone-M inherits hardware limitations of TrustZone-A and introduces more constraints.

\end{itemize}
\end{mdframed}

\begin{figure*}[hbt!]
	\begin{centering}
		\includegraphics[width=0.95\textwidth]{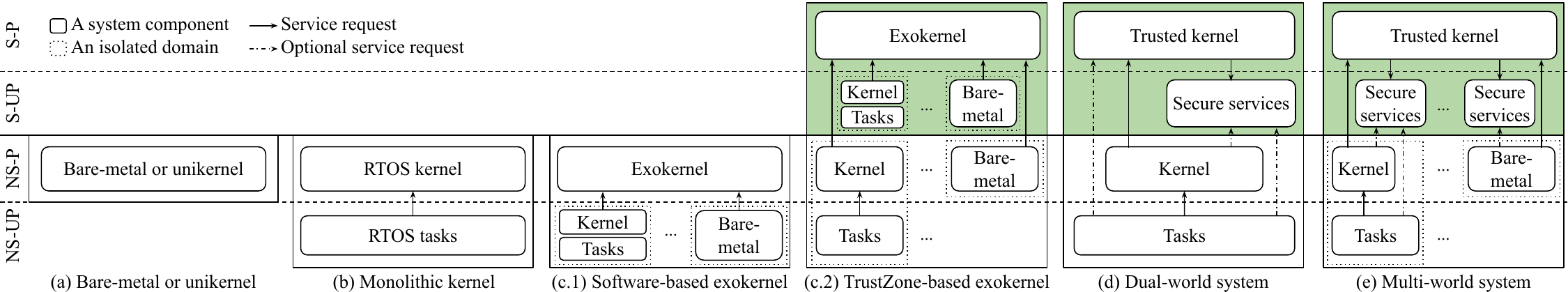}
		\vspace{-0.2cm}
		\caption{Identified Cortex-M software architectures in the collected dataset and in the literature. 
			NS-UP: non-secure unprivileged, NS-P: non-secure privileged, S-UP: secure unprivileged, S-P: secure privileged.}
		\label{fig:overview}
	\end{centering}
	\vspace{-0.5cm}
\end{figure*}

\vspace{-0.2cm}\subsection{Hardware Issues}
\label{subsec:hw_issues}

Hardware issues discuss vulnerable hardware components and hardware-supported operations. 

\vspace{.2cm}\centerline{\ul{Vendor-Agnostic Microarchitecture Issues}}


\issue{Vulnerable to microarchitectural side-channel attacks:}
Although most Cortex-M processors lack a cache or branch predictor at the microarchitectural level, there are other side channels that can leak information.

\textit{Information leakage through power analysis:}
ELMO~\cite{mccann2016elmo} demonstrates the feasibility of reversing AES S-Box output code sequences through power analysis on the Cortex-M0 processor. 
Furthermore, Vafa et al.~\cite{vafa2020efficient} successfully applied a power analysis attack to recover running instructions on the Cortex-M3 processor.

\textit{Information leakage through timing side-channels:}
MCU bus interconnect arbitration logic involves delays when multiple bus masters, such as the CPU and DMA, simultaneously access a shared secondary port, like a memory controller.
As demonstrated in BUSted~\cite{rodrigues2023busted}, the attacker can successfully bypass protections provided by the MPU and TrustZone by exploiting these timing differences.

\textit{Information leakage through long-term data remanence:}
UnTrustZone~\cite{mahmod2023untrustzone} reveals that static random-access memory (SRAM) can be manipulated to imprint and expose on-chip secrets by accelerating analog-domain changes in SRAM. 
Using this method, UnTrustZone successfully extracts AES keys and proprietary firmware from various Cortex-M devices protected by TrustZone.

\issue{Vulnerable to fault injections:}
A fault injection attack involves deliberately causing errors in a system's hardware (e.g., voltage, clock, electromagnetic) to disrupt its normal operations of a digital circuit and exploit these induced faults for malicious purposes.
Johannes Obermaier and Marc Schink et al. discussed how to escalate the debug interface permissions or execute arbitrary code by injecting faults into voltage~\cite{johannes2017firmware}, Quad-SPI bus~\cite{obermaier2020one}, and electromagnetic~\cite{schink2021security} at boot time on Cortex-M0/3/4 devices.
$\mu$-Glitch~\cite{sass2023oops} entails injecting multiple, coordinated voltage faults into Cortex-M devices to bypass the TrustZone protection, allowing leaking secrets stored in secure memory into non-secure code.

\vspace{.2cm}\centerline{\ul{Vendor-Agnostic ISA Issues}}

\issue{Fast state switch mechanism exploitable for privilege escalation:}
Cortex-M TrustZone uses the fast state switch technique to allow direct cross-state transitions from any privilege level without the need for a higher privileged secure monitor mode like Cortex-A TrustZone.
Although this feature makes cross-state transitions more efficient, it exposes vulnerabilities to a recently discovered attack known as ret2ns~\cite{ma2023dac}.
This attack leverages critical system registers and instructions used by the fast state switch to escalate privilege in the non-secure state, potentially leading to arbitrary code execution.

\issue{Improper privilege management for inter-processor debugging:}
CVE-2018-18068 reports 
that 
the debugging host's privilege level is ignored in the inter-processor debugging mode, allowing the non-secure state on both TrustZone-M and TrustZone-A to gain access to the secure state resources via the ETM~\cite{ning2019understanding, ning2021revisiting}.

\issue{Information leakage to the non-secure state due to state switches:} 
\label{issue:informationleakSW}
This could happen through memory and general-purpose and special registers:
(i) 
if a region used by the secure state is re-mapped by MPC into the non-secure state without proper sanitization, 
sensitive information will be leaked;
(ii) information leakage could happen if the general-purpose registers are not cleared when switching to the non-secure state. 
To address this issue, Arm recommends general-purpose registers that are not used to pass arguments should be cleared before state switches~\cite{Armv8MArcRef};
(iii) CVE-2021-35465 reports an issue of the floating-point lazy load multiple (\verb|VLLDM|) instruction, which allows the non-secure code to access secure state floating-point registers. 

\vspace{.2cm}\centerline{\ul{Vendor-Specific Hardware Issues}}

\issue{Improper privilege management in vendor-specific hardware features:}
Some hardware vendors introduce over-powerful hardware features that can be exploited to gain full control of the system.
For example, NXP LPC55S6x MCUs include a ROM patch controller to fix bugs in the ROM after fabrication.
CVE-2021-31532 reports that even attackers in the non-secure state and unprivileged level can utilize the ROM patch controller to reconfigure the SAU regions to gain privilege escalation.
CVE-2022-22819 shows that the ROM patch controller firmware also has a buffer overflow bug that can lead to arbitrary code execution at the privileged level.

\issue{Bypassable vendor-specific readback protection:}
Only M55 and M85 have the execute-only memory (XOM) feature, which prevents software 
or a hardware debugger from reading execute-only memory~\cite{sultan2020readback}.
For MCUs before M55, some hardware vendors implement their own hardware units to prevent reading from the debug interface, a feature known as readback protection.
For instance, the Nordic nRF51 series implements a mechanism to prevent debuggers from directly accessing flash and RAM address ranges. 
Notwithstanding, we found that only 32 out of the 1,458 Nordic samples in our dataset enable this feature. 
This protection, however, can be easily bypassed through arbitrary register read and write and single stepping in debugging~\cite{kris2015dumping}.
Though the mechanism was improved in the nRF52 series~\cite{nordicnrf52}, 
CVE-2020-27211 reports that a voltage glitch attack can still bypass it~\cite{schink2021security}.
Similar mechanisms implemented by ST~\cite{STxom}, NXP~\cite{NXPxom}, and TI~\cite{TIxom} 
are also bypassable by inferring instructions from the observed state transitions~\cite{schink2019taking}.

\begin{mdframed}[style=remarkstyle,backgroundcolor=blue!10]
\centerline{\textbf{Insights}}

\begin{itemize}[noitemsep,topsep=0pt,leftmargin=*]

\item Streamlined hardware mechanisms in Cortex-M, e.g., fast state switch, lead to new privilege management vulnerabilities and information leakage.

\item The fragmentation of the Cortex-M ecosystem has brought in new security challenges: vendors aggressively introduce over-powerful hardware, which can undermine Cortex-M systems security if not properly designed.
\end{itemize}
\end{mdframed}
 \vspace{-0.2cm}\section{Software Architectural Issues}\vspace{-0.2cm}
\label{sec:SoftIssueVul}

\subsection{Software Architectures}
\label{subsec:softarch}

As shown in Figure~\ref{fig:overview}, we identified two (i.e., a and b) software architectures in the collected firmware dataset and another three (i.e., c, d, and e) in the literature.
\textbf{\emph{Bare-metal systems and unikernels}} (a) run directly on the hardware at the highest (non-secure) privilege level.
The RTOSs in such systems are only linked as a library OS, e.g., Mbed OS bare-metal profile~\cite{MbedOS}.
We will discuss in~\ref{No or weak privilege separation:} that over 99.44\% of the 1,797 firmware belong to this category, including 66 firmware samples that use FreeRTOS and another 13 firmware use Mbed OS.
\textbf{\emph{Monolithic kernels}} (b) are the most common organization in microprocessor-based systems, e.g., Linux and Windows.  
Such systems run the kernel entirely at the privileged level, and applications run in (unprivileged) user space.
However, only 0.56\% of the firmware samples in our dataset 
fall into this category.
\textbf{\emph{Exokernels}} (c) run at the highest privilege level, virtualizing and allocating resources to RTOSs or bare-metal applications running at a lower privilege level. 
We will discuss two software-based exokernel projects, Hermes~\cite{klingensmith2018hermes} and MultiZone~\cite{pinto2020multi}, and two Cortex-M TrustZone-based exokernel projects, lLTZVisor~\cite{araujo2018lltzvisor, pinto2019virtualization} and SBIs~\cite{pansbis2022}, in \ref{Virtualization:}.
\textbf{\emph{Dual-world systems}} (d), which are enabled by TrustZone-M, run RTOSs and applications in the non-secure state, whereas secure OS/services run in the secure state.
The Trusted Firmware for Cortex-M (TF-M)~\cite{ATFM} is a reference implementation of this architecture.
\textbf{\emph{Multi-world systems}} (e) enable multiple equally-secure TEEs.
We will discuss uTango~\cite{oliveira2021utango}, one prominent example of a multi-world TEE implementation leveraging TrustZone-M in \ref{Multi-world systems:}.

\begin{mdframed}[style=remarkstyle,backgroundcolor=blue!10]
\centerline{\textbf{Insights}}

\begin{itemize}[noitemsep,topsep=0pt,leftmargin=*]

\item Despite the research progress towards more secure architectures for Cortex-M systems,
a large number of the real-world firmware in our dataset are simply bare-metal systems and unikernels.

\end{itemize}
\end{mdframed}

\subsection{Architectural Issues}
\label{subsec:sw_issues}

\begin{table*}[!t]
	\center
	\scriptsize
	\caption{Empirical Analysis of Security Features Adopted in Real-world Firmware}
	\label{tab:bin_anal}	
	\renewcommand{\arraystretch}{1}
	\setlength\tabcolsep{0.35ex}
	\rowcolors{2}{gray!15}{white}
	\begin{threeparttable}
		\begin{tabular}{l|cccc|cc|cccc|cccc|cccc|cc|cc|cc|cc}
			\hline
			\rowcolor{white}\multicolumn{1}{c|}{Hardware Vendor}                                     & \multicolumn{4}{c|}{\begin{tabular}[c]{@{}c@{}}Nordic\\ \cellcolor{white}(FirmXRay)\end{tabular}} & \multicolumn{2}{c|}{\begin{tabular}[c]{@{}c@{}}Other\\ \cellcolor{white}Nordic\end{tabular}} & \multicolumn{4}{c|}{TI}                                    & \multicolumn{4}{c|}{Telink}                                 & \multicolumn{4}{c|}{Dialog}                                & \multicolumn{2}{c|}{NXP}  & \multicolumn{2}{c|}{Cypress}  & \multicolumn{2}{c|}{ST}  & \multicolumn{2}{c}{Total} \\ \cline{2-23} \hline
			\rowcolor{white}
			Security Feature                                                                         & \multicolumn{2}{c|}{\#F}                           & \multicolumn{2}{c|}{\#D}                     & \multicolumn{2}{c|}{\#F}                                                                     & \multicolumn{2}{c|}{\#F}        & \multicolumn{2}{c|}{\#D} & \multicolumn{2}{c|}{\#F}         & \multicolumn{2}{c|}{\#D} & \multicolumn{2}{c|}{\#F}        & \multicolumn{2}{c|}{\#D} & \multicolumn{2}{c|}{\#F}  & \multicolumn{2}{c|}{\#F}      & \multicolumn{2}{c|}{\#F} & \multicolumn{2}{c}{\#F}   \\ \hline \hline
			Readback Protection~(\ref{Bypassable vendor-specific readback protection:})                         & 17          & \multicolumn{1}{c|}{2.21\%}          & 9                     & 1.75\%               & 15                                                   & 2.17\%                                & \multicolumn{2}{c|}{-}          & \multicolumn{2}{c|}{-}   & \multicolumn{2}{c|}{-}           & \multicolumn{2}{c|}{-}   & \multicolumn{2}{c|}{-}          & \multicolumn{2}{c|}{-}   & \multicolumn{2}{c|}{-}    & \multicolumn{2}{c|}{-}        & \multicolumn{2}{c|}{-}   & 32         & 1.78\%       \\
			Privilege Separation~(\ref{No or weak privilege separation:})                            & 8           & \multicolumn{1}{c|}{1.04\%}          & 5                     & 0.97\%               & 2                                                    & 0.29\%                                & 0 & \multicolumn{1}{c|}{0\%}    & 0          & 0\%         & 0  & \multicolumn{1}{c|}{0\%}    & 0        & 0\%           & 0 & \multicolumn{1}{c|}{0\%}    & 0        & 0\%           & 0           & 0\%         & 0           & 0\%             & 0           & 0\%        & 10         & 0.56\%       \\ 
			SVC for Library Call~(\ref{SVC repurposing:})                                                 & 753         & \multicolumn{1}{c|}{98.04\%}         & 500                   & 97.47\%              & 690                                                  & 100\%                                 & 2 & \multicolumn{1}{c|}{9.09\%} & 1          & 5\%         & 17 & \multicolumn{1}{c|}{8.85\%} & 17       & 14.17\%       & 0 & \multicolumn{1}{c|}{0\%}    & 0        & 0\%           & 0           & 0\%         & 2           & 2.99\%          & 2           & 50\%       & 1,466      & 81.58\%      \\
			Stack Separation~(\ref{No or weak stack separation:})                                    & 49          & \multicolumn{1}{c|}{6.38\%}          & 34                    & 6.63\%               & 82                                                   & 11.88\%                               & 0 & \multicolumn{1}{c|}{0\%}    & 0          & 0\%         & 0  & \multicolumn{1}{c|}{0\%}    & 0        & 0\%           & 3 & \multicolumn{1}{c|}{5.66\%} & 1        & 2.78\%        & 0           & 0\%         & 0           & 0\%             & 0           & 0\%        & 134        & 7.46\%       \\
			Stack Limit Register Usage~(\ref{No or weak stack separation:})            & 0           & \multicolumn{1}{c|}{0\%}             & 0                     & 0\%                  & 0                                                    & 0\%                                   & 0 & \multicolumn{1}{c|}{0\%}    & 0          & 0\%         & 0  & \multicolumn{1}{c|}{0\%}    & 0        & 0\%           & 0 & \multicolumn{1}{c|}{0\%}    & 0        & 0\%           & 0           & 0\%         & 0           & 0\%             & 0           & 0\%        & 0          & 0\%          \\
			Task Stack Ovf. Guard*~(\ref{No or weak stack separation:})                & 59          & \multicolumn{1}{c|}{96.72\%}         & 4                     & 80\%                 & 9                                                    & 32.14\%                               & \multicolumn{2}{c|}{-}          & \multicolumn{2}{c|}{-}   & \multicolumn{2}{c|}{-}           & \multicolumn{2}{c|}{-}   & \multicolumn{2}{c|}{-}          & \multicolumn{2}{c|}{-}   & \multicolumn{2}{c|}{-}    & \multicolumn{2}{c|}{-}        & \multicolumn{2}{c|}{-}   & 68         & 76.40\%      \\
			Memory Access Control (MPU)~(\ref{No or weak memory access control; executable stack:})  & 0           & \multicolumn{1}{c|}{0\%}             & 0                     & 0\%                  & 4                                                    & 0.58\%                                & 0 & \multicolumn{1}{c|}{0\%}    & 0          & 0\%         & 0  & \multicolumn{1}{c|}{0\%}    & 0        & 0\%           & 0 & \multicolumn{1}{c|}{0\%}    & 0        & 0\%           & 1           & 100\%       & 0           & 0\%             & 0           & 0\%        & 5          & 0.28\%       \\
			Memory Access Control (sMPU)~(\ref{No or weak memory access control; executable stack:}) & 19          & \multicolumn{1}{c|}{2.47\%}          & 17                    & 3.31\%               & 0                                                    & 0\%                                   & \multicolumn{2}{c|}{-}          & \multicolumn{2}{c|}{-}   & \multicolumn{2}{c|}{-}           & \multicolumn{2}{c|}{-}   & \multicolumn{2}{c|}{-}          & \multicolumn{2}{c|}{-}   & \multicolumn{2}{c|}{-}    & \multicolumn{2}{c|}{-}        & \multicolumn{2}{c|}{-}   & 19         & 1.10\%       \\
			Stack Canaries~(\ref{No or weak stack canary:})                                          & 0           & \multicolumn{1}{c|}{0\%}             & 0                     & 0\%                  & 1                                                    & 0.14\%                                & 0 & \multicolumn{1}{c|}{0\%}    & 0          & 0\%         & 0  & \multicolumn{1}{c|}{0\%}    & 0        & 0\%           & 0 & \multicolumn{1}{c|}{0\%}    & 0        & 0\%           & 0           & 0\%         & 0           & 0\%             & 0           & 0\%        & 1          & 0.06\%       \\
			Proper Instruction Sync. Barriers†~(\ref{Missing barrier instructions:})                 & 30          & \multicolumn{1}{c|}{36.59\%}         & 16                    & 27.12\%              & 68                                                   & 40\%                                  & \multicolumn{2}{c|}{-}          & \multicolumn{2}{c|}{-}   & \multicolumn{2}{c|}{-}           & \multicolumn{2}{c|}{-}   & 0 & \multicolumn{1}{c|}{0\%}    & 0        & 0\%           & \multicolumn{2}{c|}{-}    & \multicolumn{2}{c|}{-}        & \multicolumn{2}{c|}{-}   & 98         & 34.88\%       \\ \hline
		\end{tabular}
		\begin{tablenotes}
			\item[] \#F: Number of firmware, \#D: Number of devices, -: Not applicable, *: The percentage is only based on firmware that use RTOS, †: The percentage is only based on firmware that update \verb|CONTROL| with the \verb|MSR| instruction.
		\end{tablenotes}  
	\end{threeparttable}
	\vspace{-0.5cm}
\end{table*}

Software architectural issues refer to common limitations and flaws we found in real-world firmware.

\vspace{.2cm}\centerline{\ul{Lack of Privilege Management}}

\issue{No or weak privilege separation:}
As shown in Table~\ref{tab:bin_anal}, only 10 out of 1,797 samples in our dataset execute some code at the unprivileged level, and the others execute entirely at the privileged level.
Due to the lack of spatial isolation and privilege separation, 
a bug anywhere may compromise the whole system, even reverting MPU settings.

\issue{SVC repurposing:}
The \verb|SVC| instruction is designed to escalate the execution level; 
however, executing this instruction at the privileged level also transfers the control to the SVC handler. 
Surprisingly, we find that 1,466 (81.58\%) samples run everything at the privileged level and repurpose this feature to call library APIs, e.g., Nordic SoftDevice~\cite{nordicsoftdevice},
instead of privilege escalation.
The behavior is consistent across vendors, e.g., Nordic, TI, Telink, Cypress, and ST.

\vspace{.2cm}\centerline{\ul{Lack of Memory Protections}}

\issue{No or weak stack separation:}
RTOSs, such as FreeRTOS~\cite{freertosstack} and Zephyr~\cite{zephyrstack}, support multi-tasking, so each task has its own stack.
However, stack separation between the kernel and application is rarely used in bare-metal firmware.
Armv8-M also introduces stack limit registers (\texttt{PSPLIM} and \texttt{MSPLIM}) to delimit the boundaries of stacks.
However, no firmware in our dataset has been used them. 

\emph{RTOS Implementations:} 
We found that only a few RTOSs protect tasks' stacks, and only Zephyr optionally supports using stack limit registers.
When stack guard is enabled, FreeRTOS~\cite{freertoskernel} and Mbed OS~\cite{rtx5config} insert a predefined delimiter to mark the boundary of each task's stack.
Zephyr can use either \verb|PSPLIM| or an MPU-configured memory guard to prevent overwriting beyond a task's stack~\cite{zephyrstackguard}.

\emph{Empirical Analysis on Real-world Firmware:}
10 samples that adopt privilege separation (discussed in~\ref{No or weak privilege separation:}) leverage both the \verb|MSP-| and \verb|PSP-|based stacks.
In addition, another 124 samples use both the \verb|MSP-| and \verb|PSP-|based stacks without privilege separation.
All other samples (1,663; 92.54\%) only adopt a single \verb|MSP|-based stack.
59 of the 66 FreeRTOS-based firmware samples and 7 of the 13 Mbed OS-based firmware samples use task stack overflow guards.

\issue{Secure state exception stack frame manipulation:}
CVE-2020-16273 shows that the non-secure state software may manipulate the secure stacks and hijack the secure control flow
if the secure software does not properly initialize the secure stacks.
To this end, an attacker creates a fake exception return stack frame to deprivilege an interrupt.

\issue{No or weak memory access control; executable stack:}
Despite the presence of MPU, previous research suggests that it is rarely utilized in most real-world systems~\cite{clements2017EPOXY, zhou2019good, clements2018aces}.
We confirm that 1,773 of the 1,797 firmware in our dataset do not use MPU, 
which means the \emph{code}, \emph{SRAM}, and \emph{RAM} regions are executable and 
malicious code can read and write arbitrary memory. 
Out of the 24 firmware that use MPU in our dataset, five use the MPU defined by Arm. 
The remaining 19 use a vendor-specific implementation, i.e., Nordic's simplified MPU (sMPU)~\cite{nordicmpu}, which only supports
a subset of MPU features.
Specifically, sMPU only supports read and write permissions with two protection domains.


\issue{No or weak stack canary:}  
Stack canary implementation involves initializing the canary value, runtime verification, and handling mismatches.
The compiler and libraries manage the latter two, with the system initializing the canary value.
In the standard C libraries (libc),
the value of the stack canary is taken from a global variable \verb|__stack_chk_guard|. 
In modern OSs, the value of the canary is randomly initialized when a process is created. 
However, embedded systems often use a fixed canary value post-compilation or boot~\cite{tan24canary}.
Notably, there is only one \texttt{\_\_stack\_chk\_guard} for the entire physical address space.
We found that only \emph{one} of the 1,797 firmware samples in our dataset adopts it.


\issue{Missing barrier instructions:}
Barrier instructions, including data memory barrier (\verb|DMB|), data synchronization barrier (\verb|DSB|), and instruction synchronization barrier (\verb|ISB|), 
guarantee that system configurations take effect before any memory operations~\cite{membarrier}.
The omission of them is unlikely to cause any issues on most Cortex-M MCUs because they do not have out-of-order execution and branch prediction capabilities.
For MCUs that do have such capabilities, e.g., M7, this may lead to similar vulnerabilities that were discovered on microprocessors~\cite{lipp2018meltdown, kocher2018spectre, ravichandran2022pacman}.
To check if barriers are set in firmware, for any \verb|CONTROL| register update, we verify if there is an \verb|ISB| instruction in its ten subsequent instructions. 
Our analysis shows that only 98 of the 281 firmware samples (34.88\%) that update the \verb|CONTROL| register use the \verb|ISB| instruction thereafter.
However, as we cannot confirm which architecture those firmware are using, it is unclear whether the missing barrier instructions will cause issues or not.

\begin{mdframed}[style=remarkstyle,backgroundcolor=blue!10]
\centerline{\textbf{Insights}}

\begin{itemize}[noitemsep,topsep=0pt,leftmargin=*]

\item The real-world firmware samples in our dataset barely use the security features of Cortex-M and largely lack the security mitigations that are widely deployed on modern microprocessor-based systems.
\item Some software- and compiler-based mitigations, e.g., stack canaries, are less effective on MCU-based systems and should be redesigned.

\end{itemize}
\end{mdframed}
 \vspace{-0.2cm}\section{Software Implementation Issues}\vspace{-0.2cm}
\label{sec:ImpIssueVul}

\begin{table}[!t]
	\center
	\scriptsize
	\caption{Distribution of disclosed Cortex-M related CVEs (2017 - 2023)}
	\setlength\tabcolsep{0.45ex}
	\renewcommand{\arraystretch}{1}
	\rowcolors{2}{gray!15}{white}
	\label{tab:sum}
	\scalebox{0.92}{\begin{threeparttable}
			\begin{tabular}{l|cc|cc|cc|cc|cc}
				\hline
				\multicolumn{1}{c|}{HW Vendor/RTOS/Lib} & \multicolumn{2}{c|}{Critical} & \multicolumn{2}{c|}{High} & \multicolumn{2}{c|}{Medium} & \multicolumn{2}{c|}{Low} & \multicolumn{2}{c}{Total} \\ \hline \hline
				Arm                                     & 0          & 0\%              & 4         & 57.14\%       & 2          & 28.57\%        & 1        & 14.29\%       & 7         & 1.99\%        \\
				Microchip Technology                    & 1          & 14.29\%          & 2         & 28.57\%       & 4          & 57.14\%        & 0        & 0\%           & 7         & 1.99\%        \\
				Silicon Labs                            & 6          & 40.00\%          & 2         & 13.33\%       & 6          & 40.00\%        & 1        & 6.67\%        & 15        & 4.27\%        \\
				NXP Semiconductors                      & 1          & 7.69\%           & 6         & 46.15\%       & 6          & 46.15\%        & 0        & 0\%           & 13        & 3.70\%        \\
				ST Microelectronics                     & 2          & 12.50\%          & 2         & 12.50\%       & 12         & 75.00\%        & 0        & 0\%           & 16        & 4.56\%        \\
				Cypress Semiconductor                   & 0          & 0\%              & 6         & 50.00\%       & 6          & 50.00\%        & 0        & 0\%           & 12        & 3.42\%        \\
				Gigadevice                              & 0          & 0\%              & 0         & 0\%           & 6          & 100.00\%       & 0        & 0\%           & 6         & 1.71\%        \\
				Texas Instruments                       & 0          & 0\%              & 6         & 54.55\%       & 5          & 45.45\%        & 0        & 0\%           & 11        & 3.13\%        \\
				Nordic                                  & 0          & 0\%              & 2         & 50.00\%       & 2          & 50.00\%        & 0        & 0\%           & 4         & 1.14\%        \\ \hline
				Subtotal (HW vendors)                   & 10         & 10.99\%          & 30        & 32.97\%       & 49         & 53.85\%        & 2        & 2.20\%        & 91        & 25.93\%       \\ \hline \hline
				FreeRTOS                                & 3          & 15.79\%          & 9         & 47.39\%       & 7          & 36.84\%        & 0        & 0\%           & 19        & 5.41\%        \\
				CMSIS RTOS2                             & 1          & 100.00\%         & 0         & 0\%           & 0          & 0\%            & 0        & 0\%           & 1         & 0.28\%        \\
				Mbed OS                                 & 6          & 60.00\%          & 4         & 40.00\%       & 0          & 0\%            & 0        & 0\%           & 10        & 2.85\%        \\
				Zephyr                                  & 17         & 23.61\%          & 36        & 50.00\%       & 18         & 25.00\%        & 1        & 1.39\%        & 72        & 20.51\%       \\
				RIOT-OS                                 & 10         & 33.33\%          & 18        & 60.00\%       & 2          & 6.67\%         & 0        & 0\%           & 30        & 8.55\%        \\
				Contiki-ng                              & 16         & 39.02\%          & 18        & 43.90\%       & 7          & 17.07\%        & 0        & 0\%           & 41        & 11.68\%       \\
				Azure                                   & 5          & 35.71\%          & 3         & 21.43\%       & 5          & 35.71\%        & 1        & 7.14\%        & 14        & 3.99\%        \\ \hline
				Subtotal (RTOSs)                        & 58         & 31.01\%          & 88        & 47.06\%       & 39         & 20.86\%        & 2        & 1.07\%        & 187       & 53.28\%       \\ \hline \hline
				Mbed TLS                                & 6          & 20.69\%          & 12        & 41.38\%       & 11         & 37.93\%        & 0        & 0\%           & 29        & 8.26\%       \\
				WolfSSL                                 & 10         & 22.73\%          & 14        & 31.82\%       & 20         & 45.45\%        & 0        & 0\%           & 44        & 12.54\%       \\ \hline
				Subtotal (Libs)                         & 16         & 21.92\%          & 26        & 35.62\%       & 31         & 42.47\%        & 0        & 0\%           & 73        & 20.80\%       \\ \hline \hline
				Total                                   & 84         & 23.93\%          & 144       & 41.03\%       & 119        & 33.90\%        & 4        & 1.14\%        & \multicolumn{2}{c}{351}   \\ \hline
			\end{tabular}
	\end{threeparttable}}
	\vspace{-0.5cm}
\end{table}

Table~\ref{tab:sum} presents the numbers of Cortex-M related CVEs affecting nine hardware vendors, seven RTOSs, and two TLS libraries.
We break down the number based on CVSS scores~\cite{cvss}.
As shown in Table~\ref{tab:sum}, the majority of CVEs (53.85\%) affecting
hardware vendors are classified as ``medium'' severity, while the majority of CVEs affecting RTOSs (78.07\%) are categorized as either ``critical'' or ``high''.
We use a bug classification system proposed in~\cite{cerdeira2020sok} to characterize them into three major classes, i.e., validation, functional, and extrinsic.
We summarize the results in Table~\ref{tab:CVEs-con}, where
we further provide a breakdown of bugs based on the functionality and the software components. 

\begin{table*}[!ht]
	\center
	\scriptsize
	\caption{Distribution of Cortex-M software CVEs in different classes}
	\setlength\tabcolsep{2.3ex}
	\renewcommand{\arraystretch}{1}
	\rowcolors{2}{gray!15}{white}
	\label{tab:CVEs-con}	
	\begin{tabular}{l|l|l|l|l|cc}
		\hline
		\rowcolor{white}
		Bug Class         & Functions & Affected HW Vendors' SDKs  & Affected RTOSs / TLS libs &\multicolumn{2}{c}{\#Bugs}        \\ \hline \hline
		\cellcolor{white} & Communication      & \begin{tabular}[c]{@{}l@{}}\cellcolor{white}NXP (2), Microchip (5), ST (1), TI (9),\\Cypress (10), Silicon Libs (8), Nordic (3)\end{tabular} & \begin{tabular}[c]{@{}l@{}}\cellcolor{white}FreeRTOS (11), RIOT-OS (24), Mbed OS (7), Zephyr (32),\\Contiki-ng (39), Mbed TLS (14), wolfSSL (28)\end{tabular} & 193 & 57.78\%  \\
		\cellcolor{white} & Device Driver               & TF-M (1), NXP (4), ST (7)                                               & Zephyr (8), Azure (5)                                                & 25 & 7.48\%  \\
		\cellcolor{white} & Memory Allocation   & NXP  (1)                                                      & \begin{tabular}[c]{@{}l@{}}\cellcolor{white}FreeRTOS (2), RIOT-OS (2), Mbed OS (2),\\CMSIS RTOS2 (1), Zephyr (2)\end{tabular}  & 10 & 2.99\%   \\
		\cellcolor{white} & Context Switch              & TF-M (2)                                                      & FreeRTOS(1), Zephyr (3)                                                        &  6 & 1.79\%   \\
		\multirow{-7}{*}{\cellcolor{white}Validation} & Others      & Silicon Labs(5), NXP (2), Microchip (1)       & Contiki-ng (1), Zephy (10), Azure (9)   & 28 & 6.59\% \\ 
		\hline
		
		\cellcolor{white} & Protocol Implementation     & TI (1), Cypress (2), Silicon Labs (2)
		& \begin{tabular}[c]{@{}l@{}}\cellcolor{gray!15}FreeRTOS (3), RIOT-OS (4), Zephyr (13), Mbed OS (1),\\ Mbed TLS (3), wolfSSL (9)\end{tabular} &  38 & 11.38\% \\
		\cellcolor{white} & Memory Access Control      & TF-M (1), NXP (1), ST (1)   & FreeRTOS (2), Zephyr (4), Contiki-ng (1)     & 10 & 2.99\%   \\
		\multirow{-4}{*}{\cellcolor{white}Functional} & Cryptography Primitive         & TF-M (2), Microchip (1), ST (1)            &  Mbed TLS (4), wolfSSL (4)        & 12 & 3.59\%    \\ 
		\hline
		
		Extrinsic& Software Side-Channel      & ST (1) & Mbed TLS (8), wolfSSL (5)      &  14 &4.19\%    \\
		\hline
	\end{tabular}
	\vspace{-0.5cm}
\end{table*}

\vspace{-0.2cm}\subsection{Validation bugs}\vspace{-0.2cm}
Validation bugs refer to bugs that mishandle or improperly validate input and output data.
Examples are out-of-bounds read and write and improper parameter validation.
They are frequently exploited for arbitrary write and read, allowing attackers to steal/overwrite sensitive information, execute remote code, or cause a denial of service.

\issue{Validation bugs in communication components:}
Table~\ref{tab:CVEs-con}
shows that 57.78\% of validation bugs affect communication stacks, e.g., Bluetooth and TCP/IP implementations.
For instance, 
FreeRTOS has a DNS poisoning bug that does not check if a DNS answer matches an outgoing query (CVE-2018-16598).
Open-source libraries that are heavily used by Cortex-M systems, such as Mbed TLS or WolfSSL, also have 42 validation bugs.

\issue{Validation bugs in device drivers:}
Device drivers are exposed to attackers through physically-accessible peripherals, e.g., the USB interface.
We found 25 bugs that affect two hardware vendors and two RTOSs in this category.
For instance, 
the buffer overread bug of the NXP Kinetis K82 USB driver can be leveraged to access the flash (CVE-2021-44479).
The USB driver in Zephyr also has a buffer overflow bug that allows a USB-connected host to cause possible remote code execution (CVE-2020-10019). 

\issue{Validation bugs in dynamic memory allocations:}
Embedded systems commonly implement custom allocators rather than using the standard heap implementations in the Libc~\cite{gritti2022heapster}.
Bugs in heap management can result in a system crash or arbitrary code execution.
For example,
NXP's SDK, RIOT-OS, Mbed OS, and CMSIS RTOS are vulnerable to integer overflows in their allocator functions~\cite{integercve}.

\issue{Validation bugs in context switch components:}
Bugs in these components have been exploited for privilege escalation.
Zephyr uses signed integer comparison to validate the syscall number, so a negative number leads to privilege escalation (CVE-2020-10027).
TF-M has a bug allowing for out-of-bounds write in an NSC function, 
which can lead to data leakage from the secure state (CVE-2021-27562).

\issue{Validation bugs in other components:}
As discussed in~\ref{No or weak privilege separation:},
many systems execute entirely at the privileged level, and bugs in any component could lead to severe consequences.
For example, a buffer overflow in FreeRTOS' shell can cause privileged code execution (CVE-2020-10023).
Microchip's SDK has integer overflows that can be leveraged to access flash memory (CVE-2019-16127).

\vspace{-0.2cm}\subsection{Functional bugs}
Functional bugs 
refer to programming errors that do not correctly implement the intended design.

\issue{Functional bugs in protocol implementations:}
11.38\% of the functional bugs are related to protocol implementations.
For instance,
the Bluetooth controller in the Cypress SDK uses
a much shorter random number (than 128 bits) as the paring number, allowing the brute force of the random number to perform a man-in-the-middle attack during BLE pairing (CVE-2020-11957).

\issue{Functional bugs in memory access control:}
Incorrect memory access control configurations, including for MPU and TrustZone, compromise isolation.
We found eight bugs affecting one hardware vendor and two RTOSs in this category.
For example, FreeRTOS has a bug that allows any code to 
set the system privilege level (CVE-2021-43997). 

\issue{Functional bugs in cryptography primitives:}
We found four bug reports in this category.
For instance,
RIOT-OS has a nonce reuse bug in its encryption function (CVE-2021-41061) and
TF-M has a functional bug when cleaning up the memory allocated for a multi-part cryptographic operation, resulting in
a memory leak 
(CVE-2021-32032).
The implementations of PKCS \#1 v1.5 padding for RSA in the ST (CVE-2020-20949) and Microchip (CVE-2020-20950) SDKs are vulnerable to the Bleichenbacher attack~\cite{bleichenbacher1998chosen}. 
This vulnerability relies on the use of error messages or responses from the server to gain information about the validity of the padding after decryption attempts.

\vspace{-0.2cm}\subsection{Extrinsic bugs}
Extrinsic bugs refer to defects that do not belong to the validation bugs or functional errors. 

\issue{Software side-channels:}
The Lucky 13 attack in Mbed TLS (CVE-2020-16150 and CVE-2020-36423) enables an attacker to deduce secret key information by exploiting time variations in the decryption process. 
This vulnerability, specifically found in Cipher Block Chaining (CBC) mode, is based on the time differences associated with padding length.

\begin{mdframed}[style=remarkstyle,backgroundcolor=blue!10]
\centerline{\textbf{Insights}}

\begin{itemize}[noitemsep,topsep=0pt,leftmargin=*]


\item 
Most Cortex-M based production systems are written in memory-unsafe languages, e.g., C~\cite{languagesurvey}, and they suffer from memory corruption vulnerabilities.
\item Microcontrollers lack security mechanisms present in microprocessors for decades, such as privilege separation. 
Microcontroller developers may not realize the absence of features like an MMU can pose greater risks than microprocessors. Without privilege separation, any bug can be critical and compromise the entire system.


\end{itemize}
\end{mdframed}
 
 \vspace{-0.3cm}\section{Security Research}\vspace{-0.2cm}
\label{sec:projects}

\begin{figure*}[!t]
	\begin{centering}
		\includegraphics[width=.95\textwidth]{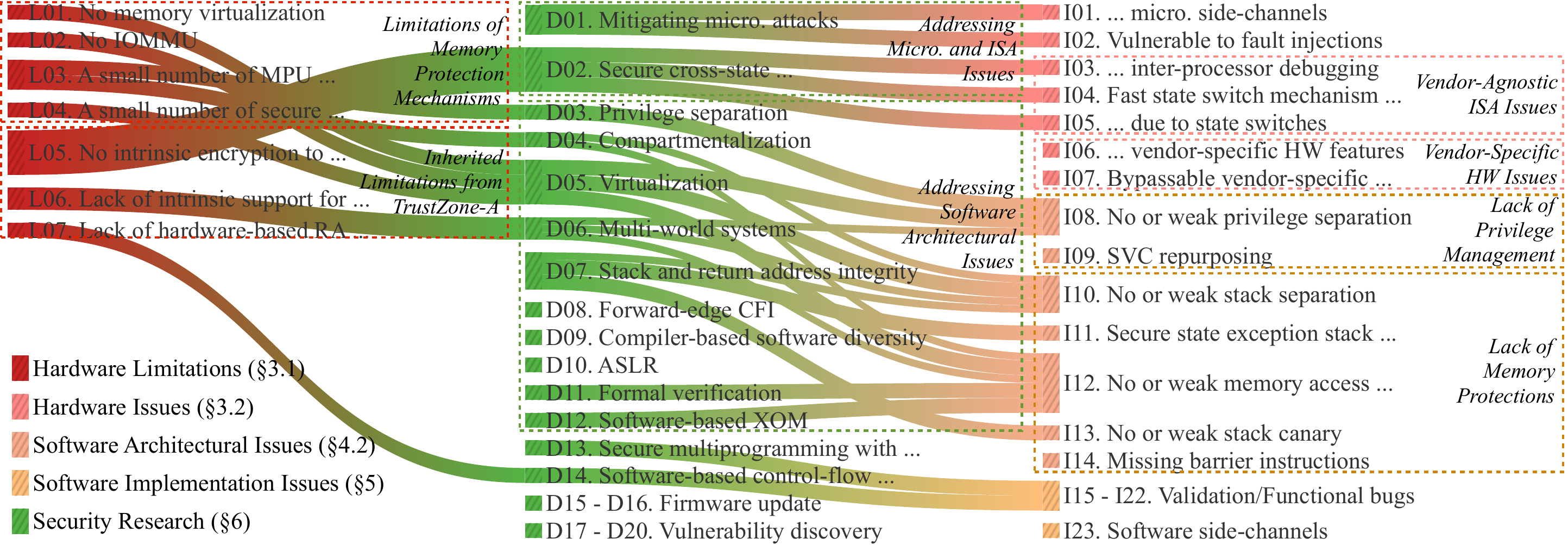}
		\vspace{-0.4cm}
		\caption{The relationships among the systematized Cortex-M related limitations, issues, and mitigations.
			The connections indicate the issues a research direction attempts to address and the limitations it needs to overcome.
			For instance, to address the issue of \emph{no or weak privilege separation}~(\ref{No or weak privilege separation:}), mitigations~(\ref{Privilege separation:}, \ref{Virtualization:}, and \ref{Multi-world systems:}) have been proposed, and they overcome some limitations (\ref{No memory virtualization:}, \ref{No input-output memory management unit:}, and \ref{A small number of MPU regions and limited sizes:}).
			An interactive version of this figure can be accessed at our anonymized repo.}
		\label{fig:id}
	\end{centering}
	\vspace{-0.5cm}
\end{figure*}

We present a taxonomy of the security research projects on Cortex-M systems. 
Figure~\ref{fig:id} depicts and summarizes the relationships among limitations, issues, and mitigations at different layers. 
Table~\ref{tab:vs-proj} presents a comparative evaluation.

\begin{table*}[!ht]
	\center
	\scriptsize
	\caption{Comparative evaluation of system isolation and attack mitigation projects for Cortex-M (\S\ref{subsec:Privilege_Separation_and_Compartment} - \S\ref{subsec:firmware-update}). The first column of the table lists the major defense mechanism proposed or adopted in a project.}
	\label{tab:vs-proj}
	\setlength\tabcolsep{0.63ex}
	\rowcolors{2}{white}{gray!15}
	\begin{threeparttable}
\begin{tabular}{l|l|c|l|ccc|cc|ccc|cc|ccc|ccccc|c|c|ccccc|cccccccccc}
			\hline
			\rowcolor{white}
			\multicolumn{1}{c}{}&\multicolumn{1}{c}{}&\multicolumn{1}{c}{}&\multicolumn{1}{c}{}&\multicolumn{3}{c}{}&\multicolumn{2}{c|}{\S\ref{subsec:hw_limits}}&\multicolumn{3}{c|}{\S\ref{subsec:hw_issues}}&\multicolumn{2}{c|}{\S\ref{subsec:sw_issues}}&\multicolumn{3}{c|}{\S\ref{sec:ImpIssueVul}}&\multicolumn{5}{c|}{}&\multicolumn{10}{c}{Self-reported average or worst case overhead (\%)} \\\hline
			\rowcolor{white}
			\multicolumn{1}{c|}{\rotatebox{90}{}}                                                                                             & \multicolumn{1}{l|}{Project}                                  & {Year} & \multicolumn{1}{l|}{Venue} & \rotatebox{90}{Input (S: source code; B: binary)} & \rotatebox{90}{Target (B: bare-metal; R: RTOS)} & \rotatebox{90}{Prototype Implementation (ISA)} & \rotatebox{90}{Limit. of Memory Prot. Mechanisms} & \rotatebox{90}{Inherited Limit. from TrustZone-A} & \rotatebox{90}{Vendor-Agnostic Micro. Issues}  & \rotatebox{90}{Vendor-Agnostic ISA Issues}& \rotatebox{90}{Vendor-specific Hardware Issues} & \rotatebox{90}{Lack of Privilege Management} & \rotatebox{90}{Lack Memory Protections} & \multicolumn{1}{c}{\rotatebox{90}{Validation Bugs}}        & \rotatebox{90}{Functional Bugs} & \rotatebox{90}{Extrinsic Bugs} & \rotatebox{90}{MPU} & \rotatebox{90}{Unprivileged Store/Load Instructions} & \rotatebox{90}{TrustZone} & \rotatebox{90}{DWT} & \rotatebox{90}{FPB} & \rotatebox{90}{Code, Binary Size Increasement} & \rotatebox{90}{Memory Overhead} & \multicolumn{1}{c}{\rotatebox{90}{Energy Consumption Overhead}} & \rotatebox{90}{Bare-metal Applications} & \multicolumn{1}{c}{\rotatebox{90}{RTOSs}} & \rotatebox{90}{BEEBS~\cite{pallister2013beebs}}   & \rotatebox{90}{CoreMark~\cite{coremark2021}} & \rotatebox{90}{CoreMark-Pro~\cite{coremarkpro2021}} & \rotatebox{90}{Dhrystone~\cite{weicker1984dhrystone}} & \rotatebox{90}{Embench~\cite{embench}} 
			\\
\hline
		\end{tabular}%
		\vspace{-0.3cm}
		\begin{tablenotes}
			\item[] 
			v7: Armv7-M, v8: Armv8-M. 
			\checkmark		: Implemented defense techniques to address at least one issue or overcome one or more limitations in the corresponding category.
			+: Need specific hardware support.
			-: Not applicable.
			\contour{black}{$\downarrow$} and \contour{black}{$\Downarrow$} represent small and big steps towards a similar goal, respectively. 
		\end{tablenotes}
	\end{threeparttable}
	\vspace{-0.5cm}
\end{table*}

\vspace{.2cm}\centerline{\ul{Addressing Hardware Issues}}\vspace{-0.4cm}

\subsection{Addressing Microarchitectural and ISA Issues}\vspace{-0.2cm}

\defense{Mitigating microarchitectural attacks:}
To mitigate information leakage through timing side-channels~(\ref{Vulnerable to microarchitectural side-channel attacks:}), BUSted~\cite{rodrigues2023busted} recommends 
disabling DMA during sensitive execution, 
and introducing random delays.
To counter information leakage through long-term data remanence, UnTrustZone~\cite{mahmod2023untrustzone} suggests initializing SRAM at startup. 
To mitigate fault injection attacks~(\ref{Vulnerable to fault injections:}), one strategy is the use of duplicate security-critical registers~\cite{barenghi2010countermeasures}. 
$\mu$Glitch suggests introducing random delays in the execution code to complicate the parameter determination process for fault injections.


\defense{Secure cross-state control and data interactions:}
One effective way to counteract privilege escalation through fast state switching (\ref{Fast state switch mechanism exploitable for privilege escalation:}) is to add additional privilege checks. 
Ret2ns~\cite{ma2023dac} suggests using address masking and MPU configuration checks to limit return targets from secure to non-secure state at the non-secure unprivileged level.
In improving privilege management for inter-processor debugging (\ref{Improper privilege management for inter-processor debugging:}), Nailgun~\cite{ning2021revisiting} employs MPU to restrict low-privilege access to debug registers.
To mitigate information leakage during cross-state switches (\ref{Information leakage to the non-secure state due to state switches:}), one approach is to implement authentication and authorization between the two states, as SeCReT~\cite{jang2015secret} does for TrustZone-A.
Secure Informer~\cite{iannillo2022ree} and ShieLD~\cite{khurshid2022shield} authenticate secure service calls from the non-secure state by verifying non-secure MPU configurations.

\vspace{.2cm}\centerline{\ul{Addressing Software Architectural Issues}}\vspace{-0.4cm}

\subsection{Separation of Privilege}\vspace{-0.1cm}
\label{subsec:Privilege_Separation_and_Compartment}

Projects in this category provide different levels of granularity in isolating and confining software modules of \emph{one} bare-metal system or \emph{one} RTOS to address \ref{No or weak privilege separation:}. 

\defense{Privilege separation:}
Solutions were proposed to automatically relegate RTOS tasks and bare-metal systems to the unprivileged level and use MPU to govern memory access. 
SAFER SLOTH~\cite{danner2014safer} dispatches tasks as interrupt handlers and lowers the privilege level in the interrupt service routine. 
EPOXY~\cite{clements2017EPOXY} automatically identifies operations requiring privileged execution (e.g., \verb|MSR|, move to system registers from general-purpose registers) in bare-metal systems.
It then relegates the whole bare-metal system to the unprivileged level and instruments privilege escalation and relegation instructions around the operations requiring privileged execution.
These privilege separation approaches only introduce a small number of context switches, introducing low overhead.

\defense{Compartmentalization:}
The projects on privilege separation~(\ref{Privilege separation:}) only split a program into privileged and unprivileged parts.
However, software modules at the same privilege level still reside in the same security and fault domain, resulting in coarse-grained memory access control~(\ref{No or weak memory access control; executable stack:}).
Several compartmentalization solutions attempt to address this issue.

\textit{Compartmentalization with heavy context switches:}
uSFI compiler~\cite{aweke2018usfi}
instruments an entry function for each module and changes cross-module procedure calls to \verb|SVC| instructions. 
ACES~\cite{clements2018aces} 
instruments binaries to enforce inter-component isolation.
MINION~\cite{kim2018securing} automatically identifies the reachable memory regions of tasks through static analysis and enforces run-time memory access control.
Because there are limited available MPU regions~(\ref{A small number of MPU regions and limited sizes:}), 
ACES and MINION propose schemes to merge the compartments.
Compared to~\ref{Privilege separation:}, compartmentalization introduces more context switches between modules; hence, the overhead is higher. 

\textit{Compartmentalization with reduced context switches:}
To reduce the overhead introduced by compartmentalization,
OPEC~\cite{zhou2022opec} leverages global variable shadowing to minimize the need for MPU regions and compartmentalizes programs to include only essential functions.
EC~\cite{khan2023ec} uses a formally verified microkernel and intra-kernel isolation to achieve compartmentalization.
CRT-C~\cite{khan2023low} compartments an RTOS into kernel, threads, and device drivers and utilizes CheckedC~\cite{elliott2018checked} to restrict their programming capabilities.

\textit{DMA-enabled compartmentalization:}
The aforementioned compartmentalization solutions do not support DMA, leaving the system vulnerable to malicious DMA-capable devices due to the absence of an IOMMU~(\ref{No input-output memory management unit:}).
D-Box~\cite{mera2022d} addresses this issue by introducing more secure MPU configurations and kernel extensions with explicit support for DMA operations.

\subsection{Virtualization and Multi-world Systems}\vspace{-0.2cm}
\label{sec:Virtualization_and_Multi-world_Systems}

Solutions in this category enable or secure \emph{multiple} bare-metal systems and RTOSs to run in an isolated fashion on \emph{one} 
MCU.

\defense{Virtualization:}
This technique can be used to support privilege separation (see~\ref{No or weak privilege separation:}).

\textit{Software-based virtualization:} 
In those solutions, bare-metal systems and RTOSs execute at the unprivileged level and the exokernel or an exception handler runs at the privileged level, as shown in Figure~\ref{fig:overview}(c.1).
A challenge is that the \verb|MSR| and \verb|MRS| (move to general-purpose registers from system registers) instructions fail silently without triggering any exceptions when executing at the unprivileged level, which can be addressed by replacing them with undefined instructions.
Examples are Hermes~\cite{klingensmith2018hermes} and MultiZone~\cite{pinto2020multi}.

\textit{TrustZone-based virtualization:} 	
As shown in Figure~\ref{fig:overview}(c.2), 
the exokernel or hypervisor runs at the highest privilege level (privileged secure state), 
and bare-metal systems and RTOSs can execute at the other three privilege levels.
Prominent examples include lLTZVisor~\cite{araujo2018lltzvisor, pinto2019virtualization} and SBIs~\cite{pansbis2022}.

\defense{Multi-world systems:}
Multiple isolation environments enhance the isolation between system components.

\textit{Real-time and secure TrustZone-assisted dual-world system:}
De facto Cortex-M TEE solutions, e.g., TF-M~\cite{ATFM}, have availability and security issues, e.g., CVE-2021-32032.
To address these issues,  
RT-TEE~\cite{wang2022rt} ensures the real-time availability of both computation and I/O
by adopting a policy-based event-driven hierarchical scheduler.
SafeTEE~\cite{schonstedt2022safetee} targets multi-core Cortex-M devices and isolates applications by assigning cores exclusively to
them. 

\textit{TrustZone-assisted multi-world system:}
As shown in Figure~\ref{fig:overview}(e),
TrustZone-assisted multi-world systems create multiple secure execution environments within the non-secure state to overcome~\ref{Lack of intrinsic support for multiple trusted execution environments:}. 
The uTango~\cite{oliveira2021utango} kernel runs in the secure state at the privileged level, while other applications, services, and OSs are isolated in their non-secure state domains.
Each domain has its own SAU configuration, which is only accessible by the uTango kernel.  

\vspace{-0.2cm}\subsection{Defeating Memory Corruption Attacks}\vspace{-0.2cm}
\label{subsec:mca}

The quest to defeat memory corruption attacks on Cortex-M systems (\ref{Validation bugs in communication components:} - \ref{Validation bugs in other components:}) largely includes adapting the security solutions for microprocessor-based systems to the resource and power constraint platforms.

\defense{Stack and return address integrity:}
Stack and return addresses are a major attack vector (\ref{No or weak stack separation:} and \ref{Secure state exception stack frame manipulation:}).
Besides stack canaries (\ref{No or weak stack canary:}), there have been many attempts to maintain stack integrity on Cortex-M.

\emph{SafeStack:} 
SafeStack~\cite{larsen2018continuing} 
keeps unsafe local variables in a separate unsafe stack while keeping the return address in the regular stack.
EPOXY implements an adapted SafeStack by (i) putting the unsafe stack on top of the RAM, (ii) making the stack grow up, and (iii) placing a region guard between the unsafe stack and other memory regions. 

\emph{Shadow stack}: 
Shadow stack~\cite{burow2019sok} stores 
protected copies of return addresses.
CaRE~\cite{nyman2017cfi} and TZmCFI~\cite{kawada2020TZmCFI} use TrustZone-M and place the shadow stack in the secure state.
To achieve low overhead, Silhouette~\cite{zhou2020silhouette} and Kage~\cite{du2022kage} restrict the writes to the shadow stack by transforming regular store instructions to unprivileged ones (\verb|STR*T|).
SUM~\cite{choi2023sum} restricts unauthorized access to the shadow stack via the MPU. 

\emph{Return address integrity}: 
$\mu$RAI~\cite{almakhdhubmurai2020} 
enforces the property of return address integrity by removing the need to spill return addresses to the stack.
Rio~\cite{kim2023rio} encrypts all return instructions in the firmware and instruments a runtime module to decrypt and execute these instructions. 
\sysnamecfi~\cite{tan2023sherloc} introduces a reconstructed call stack (RCS) approach to ensure the matching of function calls and returns.

\emph{ROP gadget removal:}
Thumb-2 instruction set~\cite{thumb2} allows the creation of ROP gadgets by jumping into the middle of a 32-bit instruction.
To replace exploitable instructions, uSFI~\cite{aweke2018usfi} and uXOM~\cite{kwon2019uxom} convert all 32-bit instructions to equivalent 16-bit instruction sequences.

\emph{Stack sealing:}
To secure the secure world stack exception frame (\ref{Secure state exception stack frame manipulation:}), Arm recommends adding an integrity signature to the bottom of the secure exception stack frame~\cite{ARMsecurestack}. 

\defense{Forward-edge control-flow integrity (CFI):}
TZmCFI adopts LLVM's forward-edge CFI~\cite{tice2014enforcing}.
CaRE calculates the absolute target addresses, stores them in a branch table, and replaces all indirect branches with \verb|SVC| instructions for run-time checking.
Silhouette and Kage insert fixed CFI labels at the beginning of every address-taken function and check the label before the jump or the function call executes.
\sysnamecfi maintains an indirect branch table
to constrain the forward target within a predetermined CFG.
InsectACIDE~\cite{wang24insect} retrieves a set of offline-computed legitimate transfer targets to validate the forward-edge transfers.

\defense{Compiler-based software diversity:}
This technique randomizes the code and data of programs~\cite{larsen2014sok} to offer weakened probabilistic protection from code reuse attacks and data corruption attacks.
However, the system memory layout remains the same after compilation.
For instance,
EPOXY~\cite{clements2017EPOXY} and Randezvous~\cite{shen2022randezvous} randomize the function order and add dummy variables to the .data and .bss regions.

\defense{Address space layout randomization (ASLR):} 
Without an MMU (\ref{No memory virtualization:}) and the dynamic loading of programs, an ASLR solution on Cortex-M needs to increase entropy and decide when to perform the randomization. 
Both HARM~\cite{shi2022harm} and fASLR~\cite{luo2022faslr} 
copy code from flash to RAM for execution and conduct randomization at the function level to increase entropy.
HARM triggers randomization periodically by SysTick exceptions,
while fASLR copies the function to a random location of RAM when it is called for the first time. 

\defense{Formal verification:}
Pip-MPU~\cite{dejon2023pip} introduces a formally verified kernel for Cortex-M.
It features user-defined, MPU-guarded multiple isolation levels and is a refactored version of the MMU-based Pip protokernel~\cite{jomaa2018formal}.
It disables exceptions and puts the kernel inside the privileged level.

\vspace{-0.2cm}\subsection{Defeating Software-based Code Disclosure}\vspace{-0.2cm}
\label{subsec:SoftCDA}

Projects in this category explore software-based XOM.
Note that these efforts cannot address \ref{Bypassable vendor-specific readback protection:}, in which a hardware debugger can disclose the contents in memory.

\defense{Software-based XOM:}
uXOM~\cite{kwon2019uxom} converts memory access instructions, excluding those that need privilege, into unprivileged ones (\verb|STR*T/LDR*T|) and sets the code region as privileged access only.
For the instructions that are not converted, uXOM instruments verification before them. 
PicoXOM~\cite{shen2020fast} implements XOM by utilizing the address range matching feature of DWT with a much lower overhead.
The DWT, however, only has up to four comparators, which limits the number of configurable XOM regions.

\vspace{.2cm}\centerline{\ul{Addressing Software Implementation Issues}}\vspace{-0.3cm}

\subsection{Memory-safe Programming}

Developing software in a manner that inherently reduces the likelihood of bugs and errors, thereby enhancing the overall safety and reliability of the system (\ref{Validation bugs in communication components:} - \ref{Software side-channels:}).

\defense{Secure multiprogramming with memory-safe languages:}
Tock~\cite{levy2017multiprogramming} takes advantage of MPU and the type-safety features of Rust to build a multiprogramming system on Cortex-M.
Rust encapsulates a large fraction of the Tock kernel with granular and type-safe interfaces. 

\subsection{Remote Attestation}\vspace{-0.1cm}

Compared to the attack mitigation discussed in~\S\ref{subsec:mca}, remote attestation
only detects adversarial presence.

\defense{Software-based control-flow and data integrity attestation:}
Control-flow attestation (CFA) extends static attestation of code to run-time control-flow paths.
DIAT~\cite{abera2019diat} provides data integrity attestation and CFA of the code that generates and processes the data.
LAPE~\cite{huo2020lape} provides a coarse-grained CFA
by grouping functions into compartments and attests 
the inter-compartment control-flow transfers.
ISC-FLAT ~\cite{neto2023isc} extends the aforementioned approaches to support interrupts, 
and ARI~\cite{wangari2023ari} formulates the property of real-time mission execution integrity.

\vspace{.2cm}\centerline{\ul{Addressing Other Issues}}\vspace{-0.3cm}

\subsection{Firmware Update}\vspace{-0.1cm}
\label{subsec:firmware-update}

\defense{Secure software update:}
ASSURED~\cite{asokan2018assured} allows a device to authenticate the source of firmware updates. 
DisPatch~\cite{kim2022reverse} 
allows end users to write patches in a domain-specific language, which DisPatch then automatically injects into the binary firmware. 
Shimware~\cite{gustafson2023shimware} investigates the challenges of updating monolithic firmware images with new security features. 
It automates 
finding safe injection locations and implementing self-checks to prevent modifications. 

\defense{Firmware hotpatching:}
While updating the whole firmware requires interrupting its normal execution (\ref{Secure software update:}), hotpatching can fix minor issues at run-time. 
HERA~\cite{niesler2021hera} uses flash patch and breakpoint (FPB) to insert hardware breakpoints and redirects the instructions at breakpoints to the patch codes on RAM.
However, FPB
is only supported on M3 and M4 MCUs.
To address this issue, RapidPatch~\cite{he2022rapidpatch} utilizes 
other hardware mechanisms, e.g., DWT.

\subsection{Vulnerability Discovery}
\label{subsec:vuldis}

\defense{Full firmware rehosting:}
One main challenge in emulating firmware on a desktop is how to model peripherals.
P$^2$IM~\cite{feng2020p} observes the MMIO access pattern of each peripheral during firmware emulation.
DICE~\cite{mera2020dice} 
improves P$^2$IM by additionally modeling DMA.
Symbolic execution that models the return value of an MMIO read as a symbolic value has also been used in firmware emulation.
Examples include Laelaps~\cite{cao2020device}, $\mu$Emu~\cite{zhou2021automatic}, Jetset~\cite{jetset}, and Fuzzware~\cite{fuzzware}.
SEmu~\cite{semu} extracts the condition-action
rules 
to dynamically synthesize peripheral models.
To sidestep the challenges in peripheral modeling,
HALucinator~\cite{clements2020halucinator}
detects and replaces hardware abstraction layer functions of major chip vendors with 
host implementations.
SAFIREFUZZ~\cite{seidel2023forming} executes embedded firmware as a Linux userspace process on systems sharing the same instruction set family as the targeted device. 
HOEDUR~\cite{scharnowski2023hoedur} employs multi-stream inputs, restructuring the traditional approach of firmware fuzzing into multiple, strictly typed, and cohesive streams, thereby enhancing mutation effectiveness and coverage.

\defense{Hardware-in-the-loop rehosting:}
Full firmware rehosting techniques
cannot accurately model more complex peripherals, such as the USB.
Hardware-in-the-loop approaches 
address this challenge by 
redirecting I/O interactions to the physical hardware.
The pioneer in this direction is Avatar~\cite{avatar},
which is followed by its
variants~\cite{avatar2,Surrogates,cor2018,HardSnap}.
Instead of redirecting I/O interactions, Frankenstein~\cite{Frankenstein} directly uses dumped firmware images from real devices
to re-establish emulator states.

\defense{On-device fuzzing:}
Existing rehosting solutions fall short in testing low-level drivers, either because they cannot provide the needed emulation fidelity
or completely sidestep driver emulation.
$\mu$AFL~\cite{li2022mu} supports on-device fuzzing with the help of a debug dongle and ETM.
Moreover, over-the-air fuzzing has been explored to find bugs in Bluetooth  controllers~\cite{SweynTooth,BRAKTOOTH}.
Lastly, to make bugs observable during fuzzing,
$\mu$SBS uses binary rewriting to instrument the firmware for sanitization checks~\cite{musbs}.
SyzTrust~\cite{wang2023syztrust} combines ETM for direct fuzzing on IoT devices with non-invasive state and code coverage tracking. 

\defense{Static methods:}
Static methods are typically geared toward detecting a particular type of bug.
For instance, PASAN~\cite{kim2021pasan} considers
concurrency issues with peripheral access.
FirmXRay~\cite{wen2020firmxray} aims to detect Bluetooth link layer vulnerabilities from bare-metal firmware.
HEAPSTER~\cite{gritti2022heapster} inspects common classes of heap vulnerabilities 
in Cortex-M monolithic firmware images.

\vspace{-0.2cm}\subsection{Other research}\vspace{-0.2cm}
\label{sec:discussion}

Solutions and ideas for other architectures may be ported to or optimized for Cortex-M with proper modifications.
For instance, the ideas of control-flow attestation (C-FLAT~\cite{abera2016c}) and operation execution integrity (OAT~\cite{sun2020oat}) apply to Cortex-M naturally but were only implemented on Cortex-A. 
In addition, on Arm Cortex-A, pointer authentication code (PAC) has been utilized to enforce spatial (e.g., return addresses~\cite{liljestrand2021pacstack} and all pointers~\cite{liljestrand2019pac}) and temporal~\cite{farkhani2021ptauth, li2022pacmem} memory safety on userspace programs and the kernel~\cite{yoo2022kernel}.

 \vspace{-0.2cm}\section{Recommendations and Future Directions}
\label{sec:rec}


\subsection{Recommendations to research community}\vspace{-0.2cm}

\rec{Explore the pros and cons of new hardware features for security:}
The hardware features of Cortex-M exhibit streamlining and differences from its Cortex-A counterparts. 
This distinction spans from the microarchitectural layer to the ISA. 
For instance, TrustZone-M is a streamlined version of TrustZone-A, and the key management for PAC~\cite{ARMpacbti} on Cortex-M significantly differs from that on Cortex-A.
All of these differences pose new challenges and opportunities in discovering their limitations and utilizing them for protections that were not possible before.


\rec{Explore diverse IoT attack models and scenarios to identify new research problems and challenges:}
The application scenarios of Cortex-M systems, e.g., (i) deployed in the field and (ii) functionality implemented in privileged mode, present unique trust models and security research opportunities, which must be addressed with extra consideration for performance, memory, and energy cost~\cite{zhao24tee, wang24insect}.
Future research should not only port the same defenses from microprocessor systems to Cortex-M systems but also address the challenges specific to MCUs.

\rec{Investigate how to facilitate the practical adoption of academic research results:}
Compared to security research on Cortex-M, its deployment significantly lags behind.
Operational research may focus on bridging the gap between security research outcomes and practical implementation. 
Such research may involve how to foster collaborations between researchers and industry practitioners, how to advocate for best practices, and how to promote educational programs to raise awareness about the importance of timely security deployment in Cortex-M systems.

\vspace{-0.2cm}\subsection{Recommendations to developers}\vspace{-0.2cm}

\rec{Securing the network communications:}
As discussed in section \S\ref{sec:ImpIssueVul}, network protocol implementations often expose many vulnerabilities including validation and functional bugs. 
This is because these protocols are designed to work with microcontroller- and microprocessor-based systems, where developers may prioritize functionalities rather than security. 
Microprocessor-based systems have advanced security mechanisms like ASLR and DEP, which can handle most security issues. However, employing vulnerable protocols on microcontroller-based systems can lead to severe problems.
Thus, microcontroller system developers should pay extra attention to security improvements, such as validating the input and output, utilizing security mechanisms discussed in section \S\ref{sec:projects}, and assessing the security of protocols before using them.

\rec{Implement privilege separation or employ RTOSs with distinct privilege levels:}
We have observed that numerous real-world firmware was built upon vendor-supplied project templates, lacking privilege separation. 
We strongly recommend developers opt for templates incorporating essential security features or, alternatively, adopt RTOSs with different privilege levels as the foundational framework for their development.

\rec{(Partially) Transition into memory-safe languages:}
A full transition into memory-safe languages, e.g., Rust, may not be immediately feasible for all Cortex-M projects due to factors like existing codebase, expertise, and project timelines~\cite{sharma2023rust}. 
Partial adoption of memory-safe languages, which provides a pragmatic and manageable approach toward embracing memory-safe languages' advantages within existing projects, can be highly valuable for enhancing the overall system robustness by mitigating memory-related issues like buffer overflows and null pointer dereferences.

\rec{Enhance the synergy between developers and the security research community:}
During our efforts to systematize security research, we noticed that some issues lack corresponding defense mechanisms (Figure~\ref{fig:id}). This could be due to incomplete publication collections, as we primarily focused on security conferences.
Nonetheless, similar to the varying levels of collaboration observed between the hacker community and academia~\cite{ndss2024keynote}, if developers and the security research community unite to share findings and insights, the security of microcontroller-based systems may be significantly improved.

 \vspace{-0.2cm}\section{Conclusion}\vspace{-0.2cm}

We present a comprehensive systematization study of the hardware and software security of Cortex-M systems. 
It covers the Cortex-M hardware architectures, security-related features, limitations, and issues. 
The study includes by far the largest empirical analysis of real-world Cortex-M firmware, characterization of reported software bugs, and an overview of state-of-the-art security research in this area. 
Based on the insights, we develop a set of recommendations for the research community and MCU software developers.
 
 \vspace{-0.2cm}\section*{Acknowledgment}\vspace{-0.2cm}
 This material is based upon work supported in part by National Science Foundation (NSF) grants (2237238, 2329704, 2207202, 2238264), a National Centers of Academic Excellence in Cybersecurity grant (H98230-22-1-0307), FCT – Fundação para a Ciência e Tecnologia within the R\&D Units Project Scope UIDB/00319/2020, and a Cisco University Research Program Fund (71858473).
 Any opinions and findings expressed in this material are those of the author(s) and do not necessarily reflect the views of United States Government or any agency thereof.

 \bibliographystyle{IEEEtran}
 \bibliography{ref}
 
 
 \appendix
 \newpage
 \section*{Appendix}

Our open-source repository contains extra information for researchers:
\begin{itemize}
	\item A Cortex-M firmware analysis tool (in the firmware\_analysis folder).
	\item A Cortex-M firmware database (in the firmware\_analysis folder).
	\item Cortex-M hardware feature test suites (in the hw\_feature\_test\_suites folder).
	\item Supplementary Material 1: Cortex-M Architecture in a Nutshell (Background.pdf).
	\item An interactive figure showcasing the relationships between Cortex-M limitations, issues, and mitigations (download relations\_interactive\_fig.html).
	\item A collection of Cortex-M-related CVEs in Google Spreadsheet.
\end{itemize}

 
 
 

 \end{document}